\documentclass[intlimits,twoside,a4paper]{article}

\usepackage{amsmath,amssymb}
\usepackage{graphicx}

\usepackage[T2A]{fontenc}
\usepackage[cp1251]{inputenc}

\usepackage{graphics}

\newcommand{\Dt}{\frac{{\rd}}{{\rd}t}}

\newcommand{\dx}{\frac{\partial}{\partial x}}
\newcommand{\dt}{\frac{\partial}{\partial t}}

\usepackage[eqsecnum]{cmpj2}

\issue{2014}{17}{4}{43002}
\doinumber{10.5488/CMP.17.43002}

\title[Chemical waves of camphor particles]%
{Exact solutions for chemical
concentration waves of self-propelling
camphor particles racing on a ring:
A novel potential dynamics perspective
}
\author[T.D. Frank]{T.D. Frank}
\address{
CESPA, Department of Psychology,
University of
Connecticut, 406 Babbidge Road, Storrs, CT 06269, USA
}

\date{Received October 8, 2014}
\authorcopyright{T.D. Frank, 2014}

\begin{document}

\maketitle

\begin{abstract}
A potential dynamics approach is developed to determine the periodic standing
and traveling wave patterns associated with self-propelling camphor objects
floating on
ring-shaped water channels. Exact solutions of the wave patterns are
derived. The bifurcation diagram describing the transition between the
immobile and
self-propelling modes of camphor objects is derived
semi-analytically. The bifurcation is of a pitchfork type which is consistent
with earlier theoretical work in which natural boundary conditions have been
considered.
\pacs
05.45.-a, 
 82.40.Ck 
\end{abstract}

\section{Introduction}

A challenge in modern-day research in biophysics and bioengineering is to
create and understand biochemical particle systems that mimic biological cell
motion. In particular,
at issue is to construct particles-surface systems in which
particles have the capability to move themselves over certain distances
by converting
chemical energy into kinetical energy~\cite{mikhailov97}. As pointed out
in reference~\cite{mikhailov97}, a fundamental class of such
man-made
self-moving systems
is given by objects that are floating on a medium, are driven by
differences in the surface tension of that medium, and at the same time
produce gradients of some substance that affects the surface
tension of the medium.
Irrespective of
engineering applications, self-propelling chemical systems allow us to study
principles that might be important for our understanding of the physics of
life, in general, and the so-called active Brownian
particles~\cite{schweitzer03book},
in particular. In fact, it has been shown that under certain conditions
self-propelling chemical ``motors'' exhibit negative friction
terms~\cite{mikhailov97,suminoyoshi08chaos} that are a
hallmark of active Brownian systems~\cite{schweitzer03book,romanczuk12epjst,frank10physletta,frank10epjb,dotov11mc,mongkolsakulvong12epjb}.
A variety of self-propelling chemical motors
have been
studied~\cite{suminoyoshi08chaos,magome96,mano05,vicario05,kitayoshi05physd,bassik08,suematsunakata10bq}.
In particular, the self-motion of
camphor particles moving on water has been extensively studied, in particular,
by Nakata and
colleagues~\cite{nakata98,hayanakata01,hayanakata02,nakata05,nakata06,suematsunakata10camphor}.
In this context, we would like to point out that not only single
camphor particles have been considered but also
the interaction between two
self-propelling camphor particles has been examined~\cite{kohira01}, the
collective motion of a small number (about 10) of camphor
particles has been
investigated~\cite{suematsunakata10pre,ikuranakata13pre}, and
the spatial and velocity distributions of many interacting
camphor particles have been determined~\cite{schulz07,soh08}.

A theoretical model for the self-motion of a single, solid camphor disc
floating on water has been
developed in terms of a Newtonian equation for the disc and a
reaction-diffusion
equation for the concentration of the surface-tension active
camphor molecules on the water
surface~\cite{naganakata04physd}. In this context,
analytical solutions of the concentration wave patterns
under natural boundary conditions have been derived and it has been shown that
the self-propelling mode bifurcates from an immobile mode by means of a
pitchfork bifurcation. However, an experimentally very
useful paradigm is the self-motion of a camphor disc on a ring
channel~\cite{nakata98,hayanakata01,hayanakata02,nakata05,nakata06,suematsunakata10camphor}.

In view of the importance of ring-shaped designs for experimental research,
the present study goes {\em
beyond the case of natural boundary conditions}. To the best of our knowledge, we will
 for the first time present a theoretical analysis
of the {\em periodic case}
 that can directly be applied to and compared with the
laboratory situation. In addition, while it is plausible to assume that the
chemical wave patterns of camphor particles on a ring exhibit a single peak,
a clear explanation why this should be the case has  not been given so far.
Such an explanation will be given below. To this end, a potential
dynamic perspective for the coupled particle-wave system will be developed.

Explicitly, the aim of the
current study is threefold.
First, we will work out a potential dynamics
approach to wave patterns of reaction-diffusion systems in order to qualitatively discuss
 the possible shapes of the camphor
concentration patterns
associated with the self-motion of camphor discs floating
on a ring channel.
Second, we will derive analytical solutions for the standing and
traveling concentration waves associated with immobile and self-propelling
camphor discs, respectively, on a ring channel. Third, we will
numerically determine the bifurcation diagram for the case of
periodic boundary conditions.
In doing so, we will show that the pitchfork bifurcation derived earlier
for natural boundary conditions can also
be found in the case of periodic boundary
conditions that is frequently used in experimental research.

We consider a camphor disc traveling on a ring channel
filled with
water. Note that the disc releases camphor to the water surface, which implies
that in this
system we need to distinguish between the camphor disc and the camphor
concentration on the water surface.
The width of the ring channel is of the disc diameter such that
the disc can move only  in one
direction, along the ring.
The ring has radius $R$ and
circumference $2\pi R$.
The
disc position is described by the periodic variable
$y(t) \in [0,2 \pi R]$, where $t$ denotes time.
Moreover, the velocity of the disc is described by
$v(t)$.
The
concentration $u$ of camphor molecules on the water surface at time $t$ and
at a particular position $x \in [0,2\pi R]$ along the ring is
described by the field variable $u(x,t)\geqslant 0$.
The dynamics of the camphor disc is given by~\cite{naganakata04physd}
\begin{subequations}
\label{eq1}
\begin{align}
\label{eq1a}
&\Dt y = v,  \\
\label{eq1b}
&\Dt v= -
\frac{a\gamma}{(a u +1)^2} \dx u\bigg|_{x=y(t)} - \mu \, v
\end{align}
\end{subequations}
with $a,\gamma,\mu>0$. Accordingly, the disc satisfies a Newtonian
equation with a force generated by the camphor concentration field $u$ and
a friction force proportional to the camphor disc velocity. From a mechanistic
point of view, the camphor concentration affects the surface tension, which in
turn acts as a force on the camphor disc (see introduction above).
The effective force given as the
first term on the right hand side of
equation~(\ref{eq1b}) can be regarded as a gradient force of a potential,
where $u$ is the
potential. That is, the camphor
disc is driven away from regions of high camphor concentrations. The
parameters $\gamma$ and $a$ are related to certain
details of the aforementioned
mechanistic relationship between camphor concentration, surface tension, and
the force acting on the camphor particle, see
reference~\cite{naganakata04physd}. Finally, in equation~(\ref{eq1}) the
parameter $\mu$ denotes the friction coefficient. The camphor concentration field $u(x,t)$ satisfies the
reaction-diffusion equation
\begin{equation}
\label{eq2}
\dt u= \frac{\partial^2}{\partial x^2} u - k \, u + F(x-y(t))
\end{equation}
with $k>0$. The term $-k \, u$ describes the decay of the camphor concentration
on the water surface due to dissolution and sublimation.
The function $F$ describes the increase of camphor concentration on the water
surface due to the camphor disc depositing camphor molecules to the water
surface. Mathematically speaking, $F$ corresponds
to a source term. In this context, note that we consider not too long time
periods during which the mass loss of the camphor disc can be
neglected.
The source term $F$ is defined by
$F=1$ for $|\Delta|\leqslant r$ and $\Delta=x-y$ and $F=0$ otherwise, where $r>0$
is the radius of the camphor disc. Note that from a mathematical point
of view for large $r$ we may not think of the disc as a circular object. We
may imagine a solid object that has the shape of a segment of the ring with
segment length $2r$ measured along the curvilinear coordinate $x$. Then,
we have $r \leqslant  \pi R$. In the case $r=\pi R$, the object would be a full,
solid ring
that covers the whole surface of the ring channel.

It is important to note that equations~(\ref{eq1}) and
(\ref{eq2}) actually
represent rescaled equations such that the variables $x$, $y$,
$u$, and even the time variable $t$ are given in dimensionless units. For
example, the (real world) laboratory time $t_\textrm{lab}$ measured in seconds
is given by the dimensionless time
$t$ occurring in equations~(\ref{eq1}) and (\ref{eq2})
divided by a rate constant measured in 1/sec that describes
the rate with which camphor is released from the camphor disc to the
water
surface~\cite{naganakata04physd}.

\section{Potential dynamics perspective of the chemical wave pattern
associated
  with a self-propelling
camphor disc}
\label{sec2}

\subsection{Potential dynamics point of view}

The objective is to study
traveling wave solutions~\cite[Sec.~7.3]{frank05book} of the form
\begin{equation}
\label{eq3}
u(x,t)=g(x-c \, t)
\end{equation}
of the model defined by equations~(\ref{eq1}) and (\ref{eq2}), where $c$
is the velocity of the traveling wave.
Let us define $z=x-c\, t$ as
the
phase coordinate of the wave.
The
reaction diffusion equation~(\ref{eq1}) describes a standing or traveling
wave pattern only if the
source $F$ depends on the phase coordinate $z$ rather than on
$x$ and $y(t)$.
Substituting  $z=x-ct$ into $\Delta=x-y(t)$, we obtain
$\Delta = z+c\,t-y(t)$. The requirement that $\Delta$ depends only on $z$
implies that $y(t)=c\,t+y_0$ with $y_0 \in [2\pi R]$. This also leads to
$v(t)=c$. The camphor disc moves with the same velocity as the wave pattern.
Substituting equation~(\ref{eq3}) into equation~(\ref{eq2}), we obtain
\begin{equation}
\label{eq4}
\frac{\partial^2}{\partial z^2}
g = - c \frac{\partial}{\partial z} g
- k \, g + F(z-y_0) \, .
\end{equation}
Likewise,
from equation~(\ref{eq1}) it follows that
\begin{equation}
\frac{\partial}{\partial z}
 g\bigg|_{z=y_0} = - \frac{\mu \, c}{a \gamma}
[a g(y_0) +1]^2 \, .
\end{equation}
The position $y_0$ of the camphor disc is arbitrary.
That is, if there is a wave pattern $g(z,y_0)$
with a
particular camphor disc position $y_0$,
then $y_0+h$ has the solution $g(z,y_0+h)=g(z-h,y_0)$,
which shifts the wave pattern by $h$.
Without loss of generality,
we put $y_0=0$ such that
\begin{equation}
\label{eq6}
\frac{{\rd}^2}{{\rd} z^2}
g = - c \frac{\partial}{\partial z} g
- k \, g + F(z)
\end{equation}
and
\begin{equation}
\label{eq7}
\frac{{\rd}}{{\rd} z}
 g\bigg|_{z=0} = - \frac{\mu \, c}{a \gamma}
[a g(0) +1]^2
\end{equation}
with $F=1$ if $z \in [0,r] \cup [2\pi R\!-\!r, 2\pi R]$
and $F=0$ otherwise.
The wave pattern is subjected to
periodic boundary conditions
$g(z)=g(z+2\pi R)$, which implies
that the relations
$g(0)=g(2\pi R)$ and
${\rd}g(0)/{\rd}z={\rd}g(2\pi R)/{\rd}z$ hold.
The traveling wave equations~(\ref{eq6}) and (\ref{eq7})
contain the standing wave equations as
special case.
For the standing wave we put $c=0$ such that
\begin{equation}
\label{eq8}
\frac{{\rd}^2}{{\rd} z^2}
g =
- k \, g + F(z)
\end{equation}
and
\begin{equation}
\label{eq9}
\frac{{\rd}}{{\rd} z}
 g\bigg|_{z=0} =0 \, .
\end{equation}
These equations have a symmetric solution $g(z)=g(-z)$.

\begin{figure}[htb]
\centerline{
\includegraphics[width=0.55\textwidth]{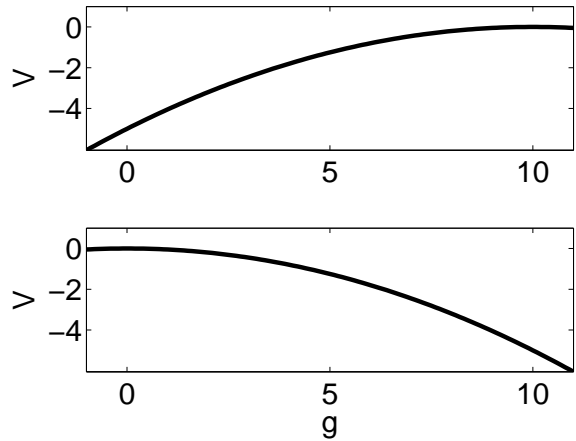}
}
\caption{Potentials of type I (top) and II (bottom) defined by
equation~(\ref{eq11}). Here $k=0.1$.}
\label{fig1}
\end{figure}

Equations~(\ref{eq6})--(\ref{eq9}) can be solved using the potential dynamics
method for solving reaction-diffusion equations (see e.g.
reference~\cite[Chap.~9]{haken77book}).
Accordingly,
$g$ is regarded as the position of a hypothetical
point particle that evolves in time $t$
and moves with a
particle velocity
$v_g$. In doing so, the phase coordinate $z$ is replaced by the
time variable $t$.
The particle motion is subjected to a potential force with a
potential
$V(g,t)$ that depends on time.
Using these replacements [i.e.,
$z \to t$ and ${\rd}g/{\rd}z \to v_g(t)$], equation~(\ref{eq6})
becomes
\begin{equation}
\label{eq10}
\Dt g = v_g  \, , \qquad
\Dt v_g = - c \, v_g - \frac{\partial}{\partial g} V(g,t)
\end{equation}
and we are looking for periodic solutions $g(t)$ with period $T=2\pi R$.
During the time intervals $[0,r]$ and $[T-r,T]$ we have $F=1$ and
the potential $V$ gives rise to the ``force''
$-{\rd}V/{\rd}g=kg-1$. By contrast, for $t \in (r,T-r)$ we have $F=0$
which implies $-{\rd}V/{\rd}g=kg$.
In total, the potential is given by
\begin{equation}
\label{eq11}
\hspace*{-3ex}
V(g,t)=
\left\{
\begin{array}{lll}
- 0.5 k (g-1/k)^2 & \text{for} & t \in \{[0,r] \cup [T-r,T]\}_{\text{modulus}\
 T}\,, \\
- 0.5 k g^2 & \text{for} & t \ \text{otherwise}\,.
\end{array}
\right.
\end{equation}
The two different potential forms described by equation~(\ref{eq11}) are
illustrated
in figure~\ref{fig1} and will be referred to as type I and II potentials,
respectively.
Both potentials are inverted parabolic potentials.
For $t \in \{[0,r] \cup [T-r,T]\}$ the potential has a peak at $g=1/k$ (top
panel),
otherwise the potential has a peak at $g=0$ (bottom panel).
Equations~(\ref{eq10}) and
(\ref{eq11}) have to be solved under
 initial conditions $g(t=0)$ and $v_g(t=0)$
that satisfy
\begin{equation}
\label{eq12}
v_g(t=0) = - \frac{\mu \, c}{a \gamma}
[a \, g(t=0) +1]^2 \ ,
\end{equation}
see equation~(\ref{eq7}).
Moreover, the periodicity condition $g(t+T)=g(t)$ implies
that the boundary conditions
$g(t=0)=g(t=T)$ and
$v_g(t=0)=v_g(t=T)$ hold. Finally, we have the constraint $g(t)\geqslant 0$ because
$g$ in the original context reflects the concentration of camphor molecules.
The potential dynamics subjected to these initial and boundary conditions
is then determined by the
two types of repulsive potentials shown in figure~\ref{fig1}.

Let us discuss period solutions $g(t)$ corresponding to
standing wave patterns $g(z)$ associated with an immobile camphor disc.
For $c=0$, equation~(\ref{eq11}) describes a
Newtonian equation without damping that is solved under the initial
 conditions $v_g=0$ and  $g(0)=A>0$. If $A>1/k$, then we have $g \to \infty$
for $t\to \infty$ because both potentials decay monotonously for $g>1/k$.
For $g(0)=1/k$, a periodic solution is not possible for $0< r<\pi R$. However,
in the limiting case $r=\pi R$, the dynamics of the
hypothetical particle is subjected
only to the potential I force with $V_\textrm{max}=1/k$
(top panel of figure~\ref{fig1})
such that $g(0)=g(t)=1/k$ is the solution of equation~(\ref{eq11}). That is, the
wave pattern $g(z)$ is a constant when the water channel is completely
filled with a solid, ring-shaped camphor object.
In summary, for  $0< r<\pi R$ we have
$g(0)=A \in (0,1/k)$. Due to the potential I force,
for $t \in [0,r]$, the particle moves
``downhill'' with respect to the type I potential $V$, that is,
to the left towards $g=0$, and reaches at time $t=r$ a point $g(r)=B$.
At $t=r$, the potential switches from type I to type II. Since
we have $v_g(r)<0$, the particle continues to move towards $g=0$. However, at
this stage it moves ``uphill'' with respect to the type II potential
$V$ and, consequently, is
de-accelerated. At a time point $t^*$, the particle has zero velocity
($v_g(t^*)=0$). At this instance, the particle has reached its minimal
position $g(t^*)=g_\textrm{min}$. Due to the impact of the
potential II force, the particle is accelerated and starts to move
``downhill'' with respect to the type II potential
$V$, that is, it moves to the right. Due to the
symmetry of the problem at hand, the time point $t^*$ is half of the period:
$t^*=T/2=\pi R$. When the particle moves to the right, $g$ increases and
eventually, at $t=T-r$, the particle reaches the position $g=B$ again.
Note that
the ``uphill'' movement $B \to g_\textrm{min}$ and the ``downhill'' movement
$g_\textrm{min} \to B$ are described by a time-reversible Newtonian equation, which
implies that the trajectory $g(t)$ is symmetric with respect to $t=T/2$.
At $t=T-r$ we have $v_g>0$. In particular, we have $v_g(T-r)=-v_g(r)$.
Moreover,
the potential switches from type II to type I.
Due to the ``initial'' velocity $v_g(T-r)>0$,
the particle goes ``uphill'' in the type I potential $V$,
slows down, and finally reaches
the location $g=A$ with $v_g=0$ at time point $t=T$.
In short, the periodic solution follows the sequence
\begin{eqnarray}
\label{eq13zus}
g(0)\!=\!g_\textrm{max}
\to g(r)\!=\!B \to g(T/2)\!=\!g_\textrm{min} \to g(T\!-\!r)\!=\!B \to g(T)\!=\!g(0)
\nonumber\\
\end{eqnarray}
and exhibits a single maximum at $g(0)$ and a single minimum at $g(T/2)$.
Moreover, the trajectory is symmetric with respect to $t=T/2$ for $t\in [0,T]$
(and symmetric with respect to $t=0$ for $t\in [-T/2,T/2]$), which means that
the periodic standing wave pattern $g(z)$ has a symmetry axis.

Periodic solutions with $v_g(0) \neq 0$ related to
traveling wave patterns
$g(z)$ induced by a self-propagating camphor disc can be discussed in a
similar way. In this context, we note that a necessary condition for a period
solution is $g(0)<1/k$ again. In order to see this,
let us assume that $g(0)>1/k$ holds. For
$c<0$, this implies $v_g(0)>0$ which implies $g \to \infty$ for $t\to \infty$
because both potentials I and II decay monotonously for $g>1/k$. For
$c>0$ and $g(0)>1/k$, we see that the particle has initial velocity
$v_g(0)<0$ and
moves towards $g=1/k$, which is the peak of the type I potential $V$
(figure~\ref{fig1}, top panel). If $|v_g(0)|$
is not sufficiently large, it will not
reach the point $g=1/k$. Rather, we will have $v_g(t')=0$ with $g(t')>1/k$
for $t' \in [0,r]$ which implies $g \to \infty$ to $t\to \infty$.
That is, a necessary condition for a periodic solution starting at $g(0)>1/k$
is that $|v_g(0)|$
is sufficiently large such that the particle passes the point $g=1/k$
during
the interval $[0,r]$. However, the particle should return somehow to
the initial position $g(0)>1/k$. Once the particle has
passed the point $g=1/k$ at $t' \in [0,r]$, the only way to return to the
subspace $g=(1/k,\infty)$ is to pass the point $g=1/k$ at a later time point
$t''>t'$ with velocity $v_g(t'')>0$. From $g(t'')=1/k$ and $v_g(t'')>0$ it
follows that $g\to\infty$ for $t\to \infty$. In summary,
periodic solutions with
$v_g(0)\neq 0$ and $g>0$ only exist for $g(0) \in (0,1/k)$.
By analogy to the previous discussion for the case $c=v_g(0)=0$,
for $c>0$ (camphor disc
and wave pattern traveling to the right on the ring coordinate $x$)
we obtain the sequence
\begin{align}
\label{eq14zus}
 g(0)&=A \to g(r)=B<A \, \to \, g(t^*)=g_\textrm{min}<B \, \to \,
g(T-r)=C>g_\textrm{min} \nonumber\\[1ex]
& \to
g(t^{**})=D=g_\textrm{max}>A \to g(T)=g(0)=A
\end{align}
with $0<r<t^*<T-r<t^{**}<T$ and $v_g=0$ at $t^*$ and $t^{**}$.
As indicated, the maximum is reached at $t^{**} \in [T-r,T]$ and does not
correspond to the initial position $g(0)$. In any case, there is a single
minimum and a single maximum, which means that a traveling wave
concentration
pattern $g(z)$ is single-peaked and exhibits a single minimum.
The condition $g(t^{**})>g(T)=g(0)$ implies for
the original problem that the camphor disc is located to the ``right''
of the peak of the traveling wave concentration pattern. For $c<0$, we conclude
in a similar vein that the traveling wave pattern is single-peaked and has a
single minimum.
As anticipated above,
the potential dynamic picture reveals that both the standing and traveling
wave patterns exhibit a single peak only. However, while the standing wave
patterns $g(z)$ exhibit a symmetry axis, traveling wave patterns $g(z)$ are
non-symmetric.

The next objective is to explicitly formulate
the scenarios described by the
sequences~(\ref{eq13zus}) and (\ref{eq14zus}) in order to
obtain analytical solutions for $g(t)$ and for the
patterns
$g(z)$.

\subsection{Standing waves}

We will derive the next symmetric solutions of equation~(\ref{eq10}) with $v_g(0)=0$.
In this case, it is sufficient to determine the trajectory $g(t)$  in the
interval $[0,T/2=\pi R]$
because, as we will see below we can take advantage of the
constraint
$v_g(T/2)=0$, see equation~(\ref{eq13zus}).
For $t \in [0,r]$, the solution $g(t)$ satisfies
equation~(\ref{eq10}) with $c=0$, $v_g(0)=0$, and $g(0) \in (0,1/k)$.
The solution reads
\begin{equation}
\label{eq13}
g(t)=\left[g(0)-\frac{1}{k}\right]\cosh(\sqrt{k}t) + \frac{1}{k} \, .
\end{equation}
In particular, we obtain
\begin{equation}
\label{eq14}
g(r) =\left[g(0)\!-\!\frac{1}{k}\right]\cosh(\sqrt{k}r) \!+\!
\frac{1}{k} \, , \qquad
v_g(r) =  \sqrt{k}\left[g(0)\!-\!\frac{1}{k}\right]\sinh(\sqrt{k}r) \, .
\end{equation}
For $t \in (r,T/2]$ the solution $g(t)$ satisfies equation~(\ref{eq10})
for $c=0$ and the initial conditions
$g(r)$ and $v_g(r)$ given above.
The solution reads
\begin{eqnarray}
\label{eq16}
g(t)&=&
\frac{1}{2}\left[g(r)+\frac{v_g(r)}{\sqrt{k}}\right]\exp\left[\sqrt{k}(t\!-\!r)\right]
+
\frac{1}{2}\left[g(r)-\frac{v_g(r)}{\sqrt{k}}\right]\exp\left[-\sqrt{k}(t\!-\!r)\right]
\nonumber\\
&=&g(r)\cosh\left[\sqrt{k}(t-r)\right]+\frac{v_g(r)}{\sqrt{k}}\sinh\left[\sqrt{k}(t-r)\right] \, ,
\end{eqnarray}
which implies that
\begin{equation}
\label{eq17}
v_g(t)=\sqrt{k}g(r)\cosh\left[\sqrt{k}(t-r)\right]+v_g(r)\cosh\left[\sqrt{k}(t-r)\right] \, .
\end{equation}
The functions~(\ref{eq13})--(\ref{eq17}) involve the unknown parameter $g(0)$.
Let us use the constraint
$v_g(T/2)=0$ to determine $g(0)$.
Substituting $t=\pi R$ with $v_g(t=\pi R)=0$
into equation~(\ref{eq17}) and substituting
equation~(\ref{eq14}) into equation~(\ref{eq17}) as well, we obtain
\begin{equation}
\label{eq18}
\xi \, w = - \frac{1}{k} \sinh\left[\sqrt{k}(\pi R-r)\right] <0
\end{equation}
with $\xi=g(0)-1/k$
and
\begin{eqnarray}
w=\sinh(\sqrt{k}r) \cosh\left[\sqrt{k}(\pi R-r)\right]+\cosh(\sqrt{k}r)
\sinh\left[\sqrt{k}(\pi R-r)\right]>0 \, . \nonumber\\
\end{eqnarray}
Since the right hand side of equation~(\ref{eq18})
is negative and $w>0$, we conclude
that $\xi<0$.
In addition, we see that
\begin{equation}
\label{eq21}
g(0)= \xi+\frac{1}{k}= \frac{1}{k} \left\{1
- \frac{\sinh\left[\sqrt{k}(\pi R-r)\right]}{w} \right\}
\end{equation}
holds.
From $w>\sinh[\sqrt{k}(\pi R -r)]$ it follows that
the factor
$\sinh[\sqrt{k}(\pi R-r)]/w>0$ is smaller than 1 such that
$g(0) \in (0,1/k)$.
That is, $g(0)$ satisfies the necessary condition for periodic solutions.

Equations~(\ref{eq13})--(\ref{eq16}) in combination with
the initial condition~(\ref{eq21})
describe the analytical solution $g(t)$ in
$t\in[0,T/2]$.
As shown above,
solutions of
equation~(\ref{eq10}) with $c=0$ and $v_g(0)=0$
are symmetric with respect to $t=0$ (and $t=T/2$). Consequently,
solutions
on a full period are given by $g(t)$ defined by
equations (\ref{eq13})--(\ref{eq16}) and (\ref{eq21}) for $t \in [0,T/2]$ and
$g(t)=g(-t)$ for $t\in [-T/2,0]$.
Likewise the pattern of the standing wave solution
$g(z)$ can be computed from
equations~(\ref{eq13})--(\ref{eq16}) and (\ref{eq21})
for $z \in [0,\pi R]$ and
$g(z)=g(-z)$ for  $z \in [-\pi R,0]$.

\begin{figure}[!t]
\centerline{
\includegraphics[width=0.55\textwidth]{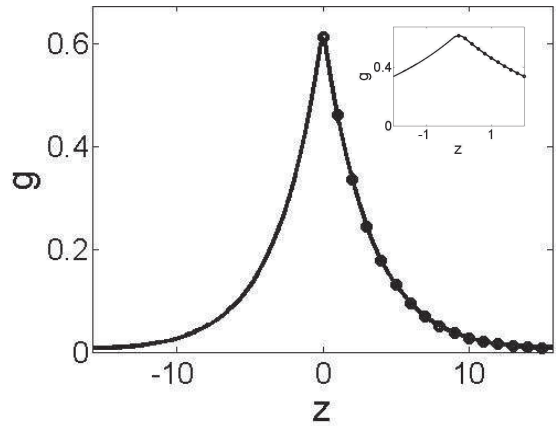}
}
\caption{Standing wave solution $g$ (solid line) computed from the
  analytical solution defined by
equations~(\ref{eq13})--(\ref{eq16}) and (\ref{eq21}). Circles correspond to
the graph $g$ determined by a numerical solution method (see text).
Insert shows a detail of the graph $g(z)$ around $z=0$.
Parameters: $k=0.1$, $R=5.0$, $r=0.2$
 (as in reference~\cite{naganakata04physd}).
}
\label{fig2}
\end{figure}

Figure~\ref{fig2}
shows $g(z)$ for a standing wave pattern.
The solid line was obtained from
equations~(\ref{eq13})--(\ref{eq16}) and (\ref{eq21})
and $g(z)=g(-z)$.
The circles were obtained by solving
equation~(\ref{eq10}) numerically
in the interval $t \in [0,T/2]$ with
initial condition $v_g(0)=0$
by means of an Euler forward
algorithm for the
dynamics of a Newtonian particle (single time step $10^{-5}$).
Numerical and analytical solution methods showed consistent results.
Although the peak of the pattern shown in figure~\ref{fig2}
looks kinky, from
equations~(\ref{eq13})--(\ref{eq16})
it follows
that
in fact the functions $g(t)$ and $g(z)$ are
smooth functions
at $t=z=0$ (i.e., continuously differentiable). This is illustrated
in the insert of figure~\ref{fig2} that depicts the standing wave pattern
in a
small region around $z=0$.

Let us briefly address
the case $r\to \pi R$. For $r=\pi R$ equation~(\ref{eq13}) yields
$g(t)=1/k$. Likewise, equation~(\ref{eq21}) reduces to $g(0)=1/k$. This is
consistent with the conclusion drawn above that for $r=\pi R$ the ring
channel exhibits a homogeneous concentration pattern $g=1/k$.

\subsection{Traveling waves}
\label{sec2p3}

Our next objective is to find solutions of equations~(\ref{eq10})--(\ref{eq12})
for $c \neq 0$ which implies $v_g(0) \neq 0$.
To this end, we will use equations~(\ref{eq10}) and (\ref{eq11})
to determine the trajectory $g(t)$ and the particle
velocity $v_g(t)$ in the full period
$[0,T]$ as function of the
unknown parameters $g(0)$, $v_g(0)$ and $c$.
We will then use the three constraints given by equation (\ref{eq12}) and
by the two periodicity requirements $g(0)=g(T)$ and $v_g(0)=v_g(T)$ to
determine
$g(0)$, $v_g(0)$ and $c$.

For $t \in [0,r]$, equation (\ref{eq10}) explicitly reads
$\rd^2g/{\rd}t^2=-c v_g + k g -1$.
The solution involves
the unknown parameters $c$, $v_g(0)$, and $g(0)$ and reads
\begin{equation}
\label{eq22}
g(t)=A_{1,0} \exp\{\lambda_1 t\} + A_{2,0} \exp\{\lambda_2 t\}+ \frac{1}{k}
\end{equation}
with eigenvalues
\begin{equation}
\label{eq23}
\lambda_1(c)=-\frac{c}{2}+\sqrt{\frac{c^2}{4}+k}>0 \,  , \qquad
\lambda_2(c)=-\frac{c}{2}-\sqrt{\frac{c^2}{4}+k}<0
\end{equation}
and amplitudes
\begin{eqnarray}
\label{eq24}
A_{1,0}=\left[g(0)-\frac{1}{k}-\frac{v_g(0)}{\lambda_2}
\right]\frac{1}{1-\lambda_1/\lambda_2} \, , \nonumber \\
A_{2,0}=\left[g(0)-\frac{1}{k}-\frac{v_g(0)}{\lambda_1}
\right]\frac{1}{1-\lambda_2/\lambda_1} \, .
\end{eqnarray}
In particular, we obtain
\begin{eqnarray}
\label{eq25}
g(r) &=& A_{1,0} \exp\{\lambda_1 r\} + A_{2,0}
\exp\{\lambda_2 r\}+ \frac{1}{k} \, , \nonumber\\
v_g(r) & = & A_{1,0} \lambda_1
\exp\{\lambda_1 r\} + A_{2,0} \lambda_2 \exp\{\lambda_2 r\} \, .
\end{eqnarray}
For $t \in (r,T-r)$, equation~(\ref{eq10})
explicitly reads ${\rd}^2g/{\rd}t^2=-c v_g + k g$. The solution involves
the initial conditions $g(r)$ and $v_g(r)$ listed above and reads
\begin{equation}
\label{eq26}
g(t)
=A_{1,r} \exp\{\lambda_1 (t-r)\} +
A_{2,r} \exp\{\lambda_2 (t-r)\}
\end{equation}
with the amplitudes
\begin{eqnarray}
\label{eq27}
A_{1,r}=\left[g(r)-\frac{v_g(r)}{\lambda_2}
\right]\frac{1}{1-\lambda_1/\lambda_2} \,  , \qquad
A_{2,r}=\left[g(r)-\frac{v_g(r)}{\lambda_1}
\right]\frac{1}{1-\lambda_2/\lambda_1} \, . \nonumber\\
\end{eqnarray}
In particular, we obtain
\begin{eqnarray}
\label{eq28}
g(T-r) &=&
A_{1,r} \exp\{\lambda_1 (T-2r)\} +
A_{2,r} \exp\{\lambda_2 (T-2r)\} \, , \nonumber\\
v_g(T-r) & = &
A_{1,r} \lambda_1 \exp\{\lambda_1 (T-2r)\} +
A_{2,r} \lambda_2 \exp\{\lambda_2 (T-2r)\} \, .
\end{eqnarray}
Finally, for $t \in [T-r,T]$
equation~(\ref{eq10})
explicitly reads ${\rd}^2g/{\rd}t^2=-c v_g + k g-1$.
The solution involves
the initial conditions $g(T-r)$ and $v_g(T-r)$ (determined above) and reads
\begin{eqnarray}
\label{eq29}
g(t)=A_{1,T-r} \exp\{\lambda_1 [t-(T-r)]\} + A_{2,T-r}
\exp\{\lambda_2 [t-(T-r)]\}
+ \frac{1}{k} \nonumber\\
\end{eqnarray}
with amplitudes
\begin{eqnarray}
\label{eq30}
A_{1,T-r}=\left[g(T-r)-\frac{1}{k}-\frac{v_g(T-r)}{\lambda_2}
\right]\frac{1}{1-\lambda_1/\lambda_2} \, , \nonumber\\
A_{2,T-r}=\left[g(T-r)-\frac{1}{k}-\frac{v_g(T-r)}{\lambda_1}
\right]\frac{1}{1-\lambda_2/\lambda_1} \, .
\end{eqnarray}
In particular, we obtain
\begin{eqnarray}
\label{eq31}
g(T) &=&
\underbrace{A_{1,T-r} \exp\{\lambda_1 r\} + A_{2,T-r}
\exp\{\lambda_2 r\}+ \frac{1}{k}}_{\displaystyle f_1}\, , \nonumber\\
v_g(T) & = &
\underbrace{A_{1,T-r} \lambda_1
\exp\{\lambda_1 r\} + A_{2,T-r} \lambda_2 \exp\{\lambda_2 r\}
}_{\displaystyle f_2} \, .
\end{eqnarray}
As mentioned above, we put
$g(0)=g(T)$ and $v_g(0)=v_g(T)$
such that
\begin{equation}
\label{eq32}
g(0)= f_1\big(g(0),v_g(0),c\big) \ , \  v_g(0)=f_2\big(g(0),v_g(0),c\big) \ .
\end{equation}
The functions $f_1$ and $f_2$ occurring in equation~(\ref{eq32})
are defined by the expressions on the
right hand sides of equation~(\ref{eq31}) and by
the amplitudes and initial conditions listed in table~\ref{tab1}.
These amplitudes and initial conditions
are explicitly defined by equations (\ref{eq24}), (\ref{eq25}),
(\ref{eq27}), (\ref{eq28}) and (\ref{eq30})
with the eigenvalues $\lambda_1$ and $\lambda_2$ given by equation (\ref{eq23}).

\begin{table}[!b]

\caption{Amplitude and ``initial conditions'' involved in the definition of the
  functions $f_1$ and $f_2$ occurring in equations (\ref{eq31}) and (\ref{eq32}).}
\label{tab1}
\vspace{2ex}
\begin{center}
\renewcommand{\arraystretch}{0}
\begin{tabular}{|c|c|c|}
\hline\hline
Amplitudes && Initial conditions\strut\\
\hline
\rule{0pt}{2pt}&&\\
\hline
$A_{1,T-r}[g(t-r),v_g(T-r),c]$
& \hspace{2ex} & $g(T-r)[A_{1,r},A_{2,r},c]$ \strut\\
$A_{2,T-r}[g(t-r),v_g(T-r),c]$ && $v_g(T-r)[A_{1,r},A_{2,r},c]$ \strut\\
\hline
$A_{1,r}[g(r),v_g(r),c]$ && $g(r)[A_{1,0},A_{2,0},c]$ \strut\\
$A_{2,r}[g(r),v_g(r),c]$ && $v_g(r)[A_{1,0},A_{2,0},c]$ \strut\\
\hline
$A_{1,0}[g(0),v_g(0),c]$ &&\strut\\
$A_{2,0}[g(0),v_g(0),c]$ &&\strut\\
\hline\hline
\end{tabular}
\renewcommand{\arraystretch}{1}
\end{center}
\end{table}

The two relations in equation~(\ref{eq32})
together with the constraint~(\ref{eq12})
provide three equations to determine the parameters
$g(0)$, $v_g(0)$, and $c$.
Once $g(0)$, $v_g(0)$, and $c$ have been determined,
the trajectory $g(t)$ can be computed from
equations~(\ref{eq22}), (\ref{eq26}) and (\ref{eq29}).
In doing so, the analytical solution for the traveling wave
pattern $g(z)$ can be found.
It can be useful to eliminate
$c$ with the help of equation~(\ref{eq12}) like
$c=-a \, \gamma \, v_g(0) /\{\mu [a g(0)+1]^2\}$.
Equation~(\ref{eq32}) then becomes
\begin{equation}
\label{eq33}
g(0)= h_1\big(g(0),v_g(0)\big) \ , \  v_g(0)=h_2\big(g(0),v_g(0)\big)
\end{equation}
with
\begin{eqnarray}
&& h_1=f_1\big(\cdot,c=-a \, \gamma \, v_g(0) /\left\{\mu [a g(0)+1]^2\right\}\big) \, , \\[1ex]
&&
h_2=f_2\big(\cdot,c=-a \, \gamma \, v_g(0) /\left\{\mu [a g(0)+1]^2\right\}\big) \, .
\end{eqnarray}

\begin{figure}[!t]
\centerline{
\includegraphics[width=0.55\textwidth]{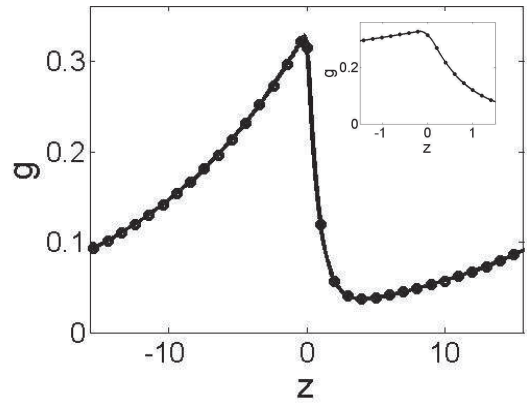}
}
\caption{Traveling wave solution $g$ (solid line) computed from the
analytical solution defined by
equations (\ref{eq22}), (\ref{eq26}) and (\ref{eq29}). Circles were obtained
by the same numerical solution method as used in figure~\ref{fig2}.
Parameters: $a=0.5$, $\gamma=2.0$, $k=0.1$, $R=5.0$, $r=0.2$, $\mu=0.1$
(as in reference~\cite{naganakata04physd}). Estimated parameters
obtained both from MATLAB \texttt{fsolve}  and
the $w_1(t)$ and $w_2(t)$ dynamical system method:
$g(0)=0.3151$, $v_g(0)=0.1516$, $c=1.1315$.
}
\label{fig3}
\end{figure}

We numerically solved
equation~(\ref{eq33})  for a given set of parameters $a$,
$\gamma$, $k$, $R$, $r$, $\mu$ in order
to obtain $g(0)$ and $v_g(0)$.
Subsequently, we calculated $c$ from
$c=-a \, \gamma \, v_g(0) /\{\mu [a g(0)+1]^2\}$.
We solved equation~(\ref{eq33}) in two different ways. First, we used the MATLAB
solver (\texttt{fsolve}) for coupled nonlinear functions. Second,
we numerically solved a dynamical
system that exhibits a fixed point consistent with equation~(\ref{eq33}).
More precisely, the two-variable dynamical system
defined by ${\rd}w_1(t)/{\rd}t=w_1-h_1(w_1,w_2)$ and
${\rd}w_2(t)/{\rd}t=h_2(w_1,w_2)-w_2$
was numerically solved (Euler forward with single time step $10^{-5}$)
for initial
conditions $w_1(0)=0.5/k$ and $w_2(0)=-0.1$ (to obtain solutions with $c>0$).
For both methods, the same numerical values for $g(0)$, $v_g(0)$, and
$c$ were obtained up to 4 digits after the decimal point.
Subsequently,
the traveling wave pattern $g(z)$ was computed from
equations~(\ref{eq22}), (\ref{eq26}) and (\ref{eq29}). The result
is shown in figure~\ref{fig3}. A detail of the
pattern for values of $z$ close to zero is shown in the insert. We see
that the camphor disc at $z=0$ is located to the ``right'' of the peak of the
concentration wave (i.e., the peak position is to the ``left'' of $z=0$).

\begin{figure}[!t]
\centerline{
\includegraphics[width=0.55\textwidth]{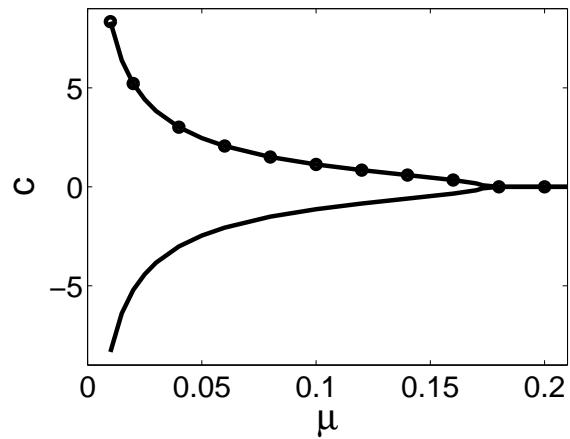}
}
\caption{Pitchfork bifurcation describing the transition of a camphor disc
  floating on a water ring from an immobile mode to a self-motion mode.
Two methods were used: MATLAB \texttt{fsolve} (solid line) and dynamical system
$w_1(t)$, $w_2(t)$ (circles). See text for details.}
\label{fig4}
\end{figure}

Finally, we compute the bifurcation diagram for the same set of parameters
$a$, $\gamma$, $k$, $R$, $r$ as used to calculate the wave pattern
$g(z)$ in figure~\ref{fig3}. However, $\mu$ was not fixed. Rather, the parameter
$\mu$ was considered as
control parameter and was varied
in the range $\mu = [0,0.2]$.
The traveling wave
velocity $c$ as function of $\mu$
was determined using the two aforementioned methods
(MATLAB \texttt{fsolve} and dynamical system $w_1(t)$, $w_2(t)$).
The propagation velocities $c$ thus
obtained are shown in figure~\ref{fig4}. The bifurcation diagram indicates that
for periodic solutions $g(z)$
there is a pitchfork bifurcation at a critical value $\mu_\textrm{c}$
such that at
$\mu=\mu_\textrm{c}$, the standing wave solution with $c=0$
becomes unstable and two stable
traveling wave
solutions emerge with either $c>0$ or $c<0$.

\section{Discussion}

A potential dynamics
approach was used to determine the shape of
chemical concentration patterns on a ring-shaped water channel
associated with a camphor disc floating
on the water surface of the channel
either in
an immobile or self-propelling fashion. The potential dynamics approach
allowed us to qualitatively determine the shape of the wave patterns and to
show that the concentration patterns are bounded from above.
In rescaled
dimensionless units, the maximal camphor
concentration of standing or traveling wave patterns
 cannot exceed the inverse of the decay constant $k$.
With the help of the potential dynamics approach,
analytical expressions for standing and traveling wave patterns
were derived. Finally, by means of the analytical expression for traveling
wave patterns, the bifurcation diagram describing the transition from
standing to traveling waves (immobility to self-motion)
when the friction parameter
$\mu$
is decreased was
determined numerically.

 The stable branches of the
bifurcation diagram for values of $\mu$
smaller than the critical value $\mu_\textrm{c}$ but close to $\mu_\textrm{c}$ (i.e.,
$\mu = \mu_\textrm{c} - \epsilon$ with $\epsilon>0$ small) may be
described like $\mu_\textrm{c}-\mu = L \, c^2 + O(c^4)$,
where $L>0$ holds and $c$ corresponds both to the
camphor disc velocity and the traveling wave
velocity. Taking the immobile-disc and standing-wave solution $c=0$
into account, this relationship between $\mu$ and
$c$ becomes $(\mu_\textrm{c}-\mu) \, c = L \, c^3 + O(c^5)$ and describes a
supercritical pitchfork bifurcation.
This relationship was analytically derived in a previous study
assuming
natural boundary conditions~\cite{naganakata04physd}. Consequently, our
considerations indicate that a supercritical
pitchfork bifurcation can also be observed in the case
of periodic boundary conditions.

In the aforementioned study it was
also shown that under natural boundary conditions
and when the
 radius of the camphor disc becomes large,
the velocity $c$ and
the control parameter $\mu$ can be related like
 $(\mu_\textrm{c}-\mu) \,
c = L_1 c^3 + L_2 c^5 + O(c^7)$ with $L_1<0$ and $L_2>0$.
When the solution $c=0$ is ignored,
 the simplified, truncated relation
$ \mu_\textrm{c}-\mu = L_1 c^2 + L_2 c^4$ describes the
 branches of self-propelling camphor discs associated with traveling wave
 concentration patterns. It has the shape of a W when
considering $\epsilon=\mu_\textrm{c}-\mu$ as a function of $c$.
Consequently, this second relationship
describes a subcritical pitchfork bifurcation when considering $c$ as a
function of $\epsilon=\mu_\textrm{c}-\mu$.
The subcritical bifurcation predicts that
for appropriately chosen values of $\mu$, there are two
different propagation velocities $c>0$ (and likewise two velocities
$c<0$). We were not able to find a counterpart of this observation for the
case of periodic boundary conditions. That is, the two numerical methods
described in section~\ref{sec2p3} to determine the bifurcation diagram did not
indicate the existence of a W shaped relationship between $\mu_\textrm{c}-\mu$ and $c$.
In this context, it is important to note that
in reference~\cite{naganakata04physd} it was shown that the range of values
$\mu$ for which two different positive (or negative) velocities
$c$ exist
is relatively small. Therefore, the issue of
a subcritical pitchfork bifurcation for
solutions of the reaction-diffusion equation~(\ref{eq2})
subjected to
periodic boundary conditions
may be addressed by
more sophisticated analytical or numerical methods that go beyond
the scope of the present study.

\clearpage

\section*{Acknowledgements}

Preparation of this manuscript was supported in part by
National Science Foundation under the INSPIRE track,
grant BCS-SBE-1344275.




\begin{thebibliography}{10}

\bibitem{mikhailov97}
Mikhailov A.S., Meink\"ohn D., In: Stochastic Dynamics, Lecture Notes in Physics, Vol.~484, {Schimansky-Geier}~L., P\"oschel~T. (Eds.), Springer, Berlin, 1997, 334--345.

\bibitem{schweitzer03book}
Schweitzer F., Brownian agents and active particles, Springer, Berlin, 2003.

\bibitem{suminoyoshi08chaos}
Sumino Y., Yoshikawa K., Chaos, 2008, \textbf{18}, 026106; \doi{10.1063/1.2943646}.

\bibitem{romanczuk12epjst}
Romanczuk P., B\"ar, Ebeling W., Lindner B., Schimansky-Geier L., Eur. Phys. J.-Spec. Top., 2012, \textbf{202}, 1; \\ \doi{10.1140/epjst/e2012-01529-y}.

\bibitem{frank10physletta}
Frank T.D., Phys. Lett. A, 2010, \textbf{374}, 3136; \doi{10.1016/j.physleta.2010.05.073}.

\bibitem{frank10epjb}
Frank T.D., Eur. Phys. J. B, 2010, \textbf{74}, 195; \doi{10.1140/epjb/e2010-00083-8}.

\bibitem{dotov11mc}
Dotov D.G., Frank T.D., Motor Control, 2011, \textbf{15}, 550.

\bibitem{mongkolsakulvong12epjb}
{Mongkolsakulvong S., Chaikhan P., Frank T.D., Eur. Phys. J. B, 2012,
  \textbf{85}, 90; \doi{10.1140/epjb/e2012-20720-4}.}

\bibitem{magome96}
Magome N., Yoshikawa K., J. Phys. Chem., 1996, \textbf{100}, 19102; \doi{10.1021/jp9616876}.

\bibitem{mano05}
Mano N., Heller A., J. Am. Chem. Soc., 2005, \textbf{127}, 11574; \doi{10.1021/ja053937e}.

\bibitem{vicario05}
Vicario J., Eelkema R., Browne W.R., Meetsma A., {La Crois} R.M., Feringa B.L.,
  Chem. Commun., 2005, 3936; \doi{10.1039/b505092h}.

\bibitem{kitayoshi05physd}
Kitahata H., Yoshikawa K., Physica D, 2005, \textbf{205}, 283; \doi{10.1016/j.physd.2004.12.012}.

\bibitem{bassik08}
Bassik N., Abebe B.T., Gracias D.H., Langmuir, 2008, \textbf{24}, 12158; \doi{10.1021/la801329g}.

\bibitem{suematsunakata10bq}
Suematsu N.J., Miyahara Y., Matsuda Y., Nakata S., J. Phys. Chem. C, 2010,
  \textbf{114}, 13340; \doi{10.1021/jp104666b}.

\bibitem{nakata98}
Nakata S., Hayashima Y., J. Chem. Soc., Faraday Trans., 1998, \textbf{94}, 3655; \doi{10.1039/a806281a}.

\bibitem{hayanakata01}
Hayashima M., Nagayama M., Nakata S., J. Phys. Chem., 2001, \textbf{105}, 5353; \doi{10.1021/jp004505n}.

\bibitem{hayanakata02}
Hayashima M., Nagayama M., Doi Y., Nakata S., Kimura M., Iida M., Phys. Chem.
  Chem. Phys., 2002, \textbf{4}, 1386; \doi{10.1039/b108686c}.

\bibitem{nakata05}
Nakata S., Matsuo K., J. Chem. Soc., Faraday Trans., 2005, \textbf{94}, 3655; \doi{10.1039/a806281a}.

\bibitem{nakata06}
Nakata S., Kirisaka J., Arima Y., Ishii T., J. Phys. Chem. B, 2006,
  \textbf{110}, 21131; \doi{10.1021/jp063827+}.

\bibitem{suematsunakata10camphor}
Suematsu N.J., Ikura Y., Nagayama M., Kitahuta H., Kawagishi N., Marukami M.,
  Nakata S., J. Phys. Chem. C, 2010, \textbf{114}, 9876; \doi{10.1021/jp101838h}.

\bibitem{kohira01}
Kohira M.I., Hayashima Y., Nagayama M., Nakata S., Langmuir, 2001, \textbf{17},
  7124; \doi{10.1021/la010388r}.

\bibitem{suematsunakata10pre}
Suematsu N.J., Nakata S., Awazu A., Nishimori H., Phys. Rev. E, 2010,
  \textbf{81},  056210;  \doi{10.1103/PhysRevE.81.056210}.

\bibitem{ikuranakata13pre}
Ikura Y.S., Heisler E., Awazu A., Nishimori H., Nakata S., Phys. Rev. E, 2013,
  \textbf{88},  012911; \\ \doi{10.1103/PhysRevE.88.012911}.

\bibitem{schulz07}
Schulz O., Markus M., J. Phys. Chem., 2007, \textbf{111}, 8175; \doi{10.1021/jp072677f}.

\bibitem{soh08}
Soh S., Bishop K.J.M., Grzybowski B.A., J. Phys. Chem. B, 2008, \textbf{112},
  10848; \doi{10.1021/jp7111457}.

\bibitem{naganakata04physd}
Nagayama B., Nakata S., Doi Y., Hayashima Y., Physica D, 2004, \textbf{194},
  151; \doi{10.1016/j.physd.2004.02.003}.

\bibitem{frank05book}
Frank T.D., Nonlinear Fokker-Planck equations: {F}undamentals and applications,
  Springer, Berlin, 2005.

\bibitem{haken77book}
Haken H., Synergetics. An introduction, Springer, Berlin, 1977.

\end{thebibliography}

\clearpage

\ukrainianpart

\title
{Точні розв'язки для хвиль хімічної концентрації самоурухомлювальних камфорних частинок, що рухаються по кільцю: нова перспектива  потенціальної динаміки
}
\author{Т.Д. Франк}
\address{
Відділ психології, Університет м. Коннектикут, CT 06269, США
}

\makeukrtitle

\begin{abstract}
\tolerance=3000%
Розвинуто метод потенціальної динаміки для того, щоб вивчити  картини періодичних стоячих і біжучих хвиль, пов'язаних із
самоурухомлювальними камфорними об'єктами, які рухаються на кільцеподібних водяних каналах.
Отримано точні розв'язки  хвильвових картин. Отримано напіваналітично діаграму біфуркації, яка описує перехід між нерухомою і
самоурухомлювальною модами камфорних об'єктів. Біфуркація є вилоподібна, що узгоджується з попередньою теоретичною роботою, в якій
розглянуто природні граничні умови.
\keywords хвилі хімічної концентрації, саморух, вилоподібна біфуркація
\end{abstract}

\end{document}